\documentclass[preprint,prd,nofootinbib,tightenlines]{revtex4}
\usepackage{epsf}
\begin{document}
\hskip 1cm
\vskip .1 cm
\title{\Large  Inflation and String Cosmology\footnote{This is an extended version of my talks at the  SLAC Summer School ``Nature's Greatest Puzzles," at the conference Cosmo04 in Toronto, at the VI Mexican School on Gravitation, and at the XXII Texas Symposium on Relativistic Astrophysics in 2004.}}

\vskip 1cm
\
 
 \
 
\author{Andrei Linde}
\affiliation{Department of Physics, Stanford University, Stanford, CA 94305,
USA}

\

\begin{abstract}
After  25 years of its existence, inflationary theory
gradually becomes  the standard cosmological paradigm. However, we
still do not know which of the many versions of inflationary
cosmology will be favored by the future observational data.
Moreover, it may be quite nontrivial to obtain a natural
realization of inflationary theory in the context of the ever
changing theory of all fundamental interactions.
In this paper I will describe the history and the present  status of inflationary cosmology, including recent attempts to implement inflation in the context of string theory.
\end{abstract}
\pacs{98.80.Cq \hskip 3.3cm SU-ITP-05/11 \hskip 3.3cm \ hep-th/0503195}
 \maketitle

\tableofcontents 

\newpage

\section {Brief history of inflation}

The first model of inflationary type was proposed by Alexei
Starobinsky \cite{Star}. It was based on investigation of
conformal anomaly in quantum gravity.  This model was rather complicated, and its goal was somewhat different from the goals of inflationary cosmology. Instead of attempting to solve the homogeneity and isotropy problems, Starobinsky considered the model of  the universe which was homogeneous and isotropic from the very beginning, and emphasized that his scenario was ``the extreme opposite of Misner's initial ``chaos''.'' 

On the other hand, Starobinsky model  did not suffer from the graceful exit problem, and it was the first model predicting  gravitational waves with flat spectrum \cite{Star}. More importantly, the first mechanism of production of adiabatic perturbations of metric with a flat spectrum, which are responsible for galaxy production, and which were found by the observations of the CMB anisotropy,  was proposed by  Mukhanov  and Chibisov \cite{Mukh} in the context of this model.

A much simpler inflationary model with a very clear physical
motivation was proposed by Alan Guth \cite{Guth}.  His model,
which is now called ``old inflation,'' was based on the theory of
supercooling during the cosmological phase transitions
\cite{Kirzhnits}. Even though this scenario did not work,  it
played a profound role in the development of inflationary
cosmology since it contained a very clear explanation how
inflation may solve the major cosmological problems.

According to this scenario,  inflation is as   exponential
expansion of the universe in a supercooled false vacuum state.
False vacuum is a metastable state without any fields or particles
but with large energy density. Imagine a universe filled with such
``heavy nothing.'' When the universe expands, empty space remains
empty, so its energy density does not change. The universe with a
constant energy density expands exponentially, thus we have
inflation in the false vacuum. This expansion makes the universe very big and very flat. Then the false vacuum decays, the
bubbles of the new phase collide, and our universe becomes hot.

Unfortunately, this simple and intuitive picture of inflation in the false vacuum state, which is often presented in the popular literature, is somewhat misleading. If the bubbles of the new phase are formed near
each other, their collisions make the universe extremely
inhomogeneous. If they are formed far away from each other, each
of them represents a separate open universe with a vanishingly
small $\Omega$. Both options are unacceptable, which has lead to the conclusion that this scenario cannot be improved (graceful exit problem) \cite{Guth,Hawking:1982ga,Guth:1982pn}.

This problem  was resolved with the invention of the new inflationary theory \cite{New,New2}. In this theory, inflation may
begin either in the false vacuum,  or in an unstable state at the
top of the effective potential. Then the inflaton field $\phi$
slowly rolls down to the minimum of its effective potential.  The
motion of the field away from the false vacuum is of crucial
importance: density perturbations produced during the slow-roll
inflation are inversely proportional to $\dot \phi$
\cite{Mukh,Hawk,Mukh2}. Thus the key difference between the new
inflationary scenario and the old one is that the useful part of
inflation in the new scenario, which is responsible for the
homogeneity of our universe, does {\it not} occur in the false
vacuum state, where $\dot\phi =0$. 

Some authors recently started using a generalized notion of the false vacuum, defining it as a vacuum-like state with a slowly changing energy density. While the difference between this definition and the standard one is very subtle, it is exactly this difference that is responsible for solving all problems of the old inflation scenario, and it is exactly this subtlety that made it so difficult to find this solution. 

The new inflationary scenario became so popular in the beginning of the 80's that even now most textbooks on astrophysics incorrectly describe inflation as an exponential expansion during high temperature phase transitions in
grand unified theories. Unfortunately,  new inflation 
was plagued by its own problems. It works only if the effective
potential of the field $\phi$ has a very a flat plateau near $\phi
= 0$, which is somewhat artificial. In most versions of this
scenario the inflaton field has an extremely small coupling
constant, so it could not be in thermal equilibrium with other
matter fields. The theory of cosmological phase transitions, which
was the basis for old and new inflation, did not work in such a
situation. Moreover, thermal equilibrium requires many particles
interacting with each other. This means that new inflation could
explain why our universe was so large only if it was very large
and contained many particles from the very beginning. Finally,
inflation in this theory begins very late. During the preceding
epoch the universe can easily collapse or become so inhomogeneous
that inflation may never happen \cite{book}.

Old and new inflation represented a substantial but incomplete
modification of the big bang theory. It was still assumed that the
universe was in a state of thermal equilibrium from the very
beginning, that it was relatively homogeneous and large enough to
survive until the beginning of inflation, and that the stage of
inflation was just an intermediate stage of the evolution of the
universe. In the beginning of the 80's these assumptions seemed
most natural and practically unavoidable. On the basis of all
available observations (CMB, abundance of light elements)
everybody believed that the universe was created in a hot big
bang. That is why it was so difficult to overcome a certain
psychological barrier and abandon all of these assumptions. This
was done with the invention of the chaotic inflation scenario
\cite{Chaot}. This scenario resolved all problems of old and new
inflation. According to this scenario, inflation may occur even in
the theories with simplest potentials such as $V(\phi) \sim
\phi^n$. Inflation may begin even if there was no thermal
equilibrium in the early universe, and it may start even at the
Planckian density, in which case the problem of initial conditions
for inflation can be easily resolved \cite{book}.

\section{Chaotic Inflation}

Consider  the simplest model of a scalar field $\phi$ with a mass
$m$ and with the potential energy density $V(\phi)  = {m^2\over 2}
\phi^2$. Since this function has a minimum at $\phi = 0$,  one may
expect that the scalar field $\phi$ should oscillate near this
minimum. This is indeed the case if the universe does not expand,
in which case equation of motion for the scalar field  coincides
with equation for harmonic oscillator, $\ddot\phi = -m^2\phi$.

However, because of the expansion of the universe with Hubble
constant $H = \dot a/a $, an additional  term $3H\dot\phi$ appears
in the harmonic oscillator equation:
\begin{equation}\label{1x}
 \ddot\phi + 3H\dot\phi = -m^2\phi \ .
\end{equation}
The term $3H\dot\phi$ can be interpreted as a friction term. The
Einstein equation for a homogeneous universe containing scalar
field $\phi$ looks as follows:
\begin{equation}\label{2x}
H^2 +{k\over a^2} ={1\over 6}\, \left(\dot \phi ^2+m^2 \phi^2)
\right) \ .
\end{equation}
Here $k = -1, 0, 1$ for an open, flat or closed universe
respectively. We work in units $M_p^{-2} = 8\pi G = 1$.

If   the scalar field $\phi$  initially was large,   the Hubble
parameter $H$ was large too, according to the second equation.
This means that the friction term $3H\dot\phi$ was very large, and
therefore    the scalar field was moving   very slowly, as a ball
in a viscous liquid. Therefore at this stage the energy density of
the scalar field, unlike the  density of ordinary matter, remained
almost constant, and expansion of the universe continued with a
much greater speed than in the old cosmological theory. Due to the
rapid growth of the scale of the universe and a slow motion of the
field $\phi$, soon after the beginning of this regime one has
$\ddot\phi \ll 3H\dot\phi$, $H^2 \gg {k\over a^2}$, $ \dot \phi
^2\ll m^2\phi^2$, so  the system of equations can be simplified:
\begin{equation}\label{E04}
H= {\dot a \over a}   ={ m\phi\over \sqrt 6}\ , ~~~~~~  \dot\phi =
-m\  \sqrt{2\over 3}     .
\end{equation}
The first equation shows that if the field $\phi$ changes slowly,
the size of the universe in this regime grows approximately as
$e^{Ht}$, where $H = {m\phi\over\sqrt 6}$. This is the stage of
inflation, which ends when the field $\phi$ becomes much smaller
than $M_p=1$. Solution of these equations shows that after a long
stage of inflation  the universe initially filled with the field
$\phi   \gg 1$  grows  exponentially \cite{book}, 
\begin{equation}\label{E04aa}
 a= a_0
\ e^{\phi^2/4} \   .
\end{equation}

This is as simple as it could be. Inflation does not require
supercooling and tunneling from the false  vacuum \cite{Guth}, or
rolling from an artificially flat top of the effective potential
\cite{New,New2}. It appears in the theories that can be as simple as a
theory of a harmonic oscillator \cite{Chaot}. Only when it was
realized, I started to really believe that inflation is not just a trick
necessary to fix  problems of the old big bang theory, but a
generic cosmological regime.

\begin{figure}
\leavevmode\epsfysize=7 cm \epsfbox{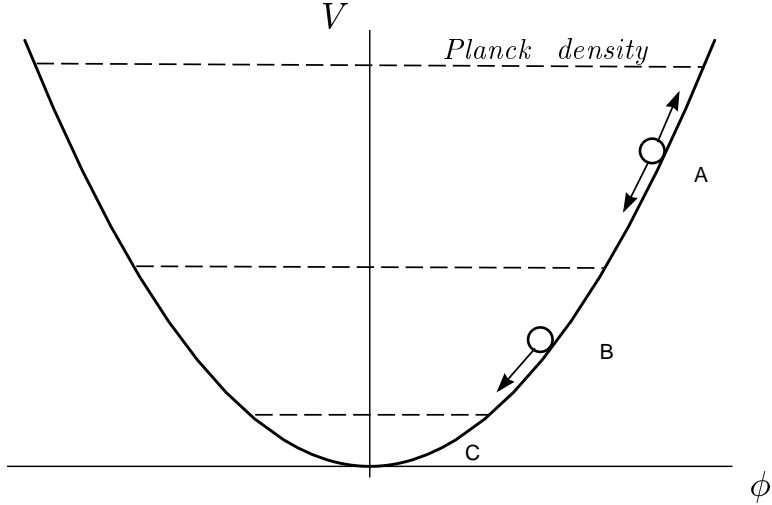}

\caption{Motion of the scalar field in the theory with $V(\phi) =
{m^2\over 2} \phi^2$. Several different regimes are possible,  depending
on the value of the field $\phi$. If the potential energy density of the
field is greater than the Planck density $M_p^4$,
quantum fluctuations of space-time are so strong that one cannot describe
it in usual terms. Such a state   is called space-time foam. At a
somewhat smaller energy density  (region A: $m M_p^3 < V(\phi) < M_p^4$)
quantum fluctuations of space-time are small, but quantum fluctuations of
the scalar field $\phi$ may be large. Jumps of the scalar field due to
quantum fluctuations lead to a process of eternal self-reproduction of
inflationary universe which we are going to discuss later. At even
smaller values of $V(\phi)$ (region B: $m^2  M_p^2 < V(\phi) < m M_p^3$ )
fluctuations of the field $\phi$ are small; it slowly moves down as a
ball in a viscous liquid. Inflation occurs both in the region A and
region B. Finally, near the minimum of $V(\phi)$ (region C) the scalar
field rapidly oscillates, creates pairs of elementary particles, and the
universe becomes hot.} \label{fig:Fig1}
\end{figure}

But what's about the initial conditions required for chaotic inflation?
Let us consider first a closed Universe of initial  size $l \sim 1$
(in Planck units), which emerges  from the
space-time foam, or from singularity, or from `nothing'  in a state
with  the Planck density $\rho \sim 1$. Only starting from this moment, i.e. at $\rho
\lesssim 1$, can we describe this domain as  a {\it classical} Universe.  Thus,
at this initial moment the sum of the kinetic energy density, gradient energy
density, and the potential energy density  is of the order unity:\, ${1\over
2} \dot\phi^2 + {1\over 2} (\partial_i\phi)^2 +V(\phi) \sim 1$.

We wish to emphasize, that there are no {\it a priori} constraints on
the initial value of the scalar field in this domain, except for the
constraint ${1\over 2} \dot\phi^2 + {1\over 2} (\partial_i\phi)^2 +V(\phi) \sim
1$.  Let us consider for a moment a theory with $V(\phi) = const$. This theory
is invariant under the {\it shift symmetry}  $\phi\to \phi + c$. Therefore, in such a
theory {\it all} initial values of the homogeneous component of the scalar field
$\phi$ are equally probable.  

The only constraint on the  amplitude of the field appears if the
effective potential is not constant, but grows and becomes greater
than the Planck density at $\phi > \phi_p$, where  $V(\phi_p) = 1$. This
constraint implies that $\phi \lesssim \phi_p$,
but there is no reason to expect that initially
$\phi$ must be much smaller than $\phi_p$. This suggests that the  typical initial value
 of the field $\phi$ in such a  theory is  $\phi 
\sim  \phi_p$.

Thus, we expect that typical initial conditions correspond to
${1\over 2}
\dot\phi^2 \sim {1\over 2} (\partial_i\phi)^2\sim V(\phi) = O(1)$.
If ${1\over 2} \dot\phi^2 + {1\over 2} (\partial_i\phi)^2
 \lesssim V(\phi)$
in the domain under consideration, then inflation begins,
and then within
the Planck time the terms  ${1\over 2} \dot\phi^2$ and ${1\over 2}
(\partial_i\phi)^2$ become much smaller than $V(\phi)$, which ensures
continuation of inflation.  It seems therefore that chaotic inflation
occurs under rather natural initial conditions, if it can begin at $V(\phi)
\sim 1$. 

As we will see shortly, the realistic value of the mass $m$ is about $3\times 10^{-6}$, in Planck units. Therefore, according to Eq. (\ref{E04aa}), the total amount of inflation achieved starting from $V(\phi) \sim 1$ is of the order $10^{10^{10}}$. The total duration of
inflation in this model is about $10^{-30}$ seconds. When inflation
ends, the scalar field $\phi$ begins to   oscillate near the
minimum of $V(\phi)$. As any rapidly oscillating classical field,
it looses its energy by creating pairs of elementary particles.
These particles interact with each other and come to a state of
thermal equilibrium with some temperature $T_{r}$ \cite{oldtheory,KLS,tach,Desroche:2005yt,latticeold,latticeeasy,thermalization}.
From this time on, the universe can be described by the usual big
bang theory.

The main difference between inflationary theory and the old
cosmology becomes clear when one calculates the size of a typical
inflationary domain at the end of inflation. Investigation of this
question    shows that even if  the initial size of   inflationary
universe  was as small as the Planck size $l_P \sim 10^{-33}$ cm,
after $10^{-30}$ seconds of inflation   the universe acquires a
huge size of   $l \sim 10^{10^{10}}$ cm! This number is
model-dependent, but in all realistic models the  size of the
universe after inflation appears to be many orders of magnitude
greater than the size of the part of the universe which we can see
now, $l \sim 10^{28}$ cm. This immediately solves most of the
problems of the old cosmological theory \cite{Chaot,book}.

Our universe is almost exactly homogeneous on  large scale because
all inhomogeneities were exponentially stretched during inflation.
The density of  primordial monopoles  and other undesirable
``defects'' becomes exponentially diluted by inflation.   The
universe   becomes enormously large. Even if it was a closed
universe of a size
 $\sim 10^{-33}$ cm, after inflation the distance between its ``South'' and
``North'' poles becomes many orders of magnitude greater than
$10^{28}$ cm. We see only a tiny part of the huge cosmic balloon.
That is why nobody  has ever seen how parallel lines cross. That
is why the universe looks so flat.

If our universe initially consisted of many domains with
chaotically distributed scalar field  $\phi$ (or if one considers
different universes with different values of the field), then
domains in which the scalar field was too small never inflated.
The main contribution to the total volume of the universe will be
given by those domains which originally contained large scalar
field $\phi$. Inflation of such domains creates huge homogeneous
islands out of initial chaos. (That is why I called this scenario
``chaotic inflation.'') Each  homogeneous domain in this scenario
is much greater than the size of the observable part of the
universe.

The first models of chaotic inflation were based on the theories
with polynomial potentials, such as $V(\phi) = \pm {m^2\over 2}
\phi^2 +{\lambda\over 4} \phi^4$. But, as was emphasized in  \cite{Chaot}, the main idea of this scenario is quite generic. One should consider any particular
potential $V(\phi)$, polynomial or not, with or without
spontaneous symmetry breaking, and study all possible initial
conditions without assuming that the universe was in a state of
thermal equilibrium, and that the field $\phi$ was in the minimum
of its effective potential from the very beginning.

This scenario strongly deviated from the standard lore of the hot
big bang theory and was psychologically difficult to accept.
Therefore during the first few years after invention of chaotic
inflation many authors claimed that the idea of chaotic initial
conditions is unnatural, and made attempts to realize the new
inflation scenario based on the theory of high-temperature phase
transitions, despite numerous problems associated with it. Some
authors believed that the theory must satisfy so-called `thermal
constraints' which were necessary to ensure that the minimum of
the effective potential at large $T$ should be at $\phi=0$
\cite{OvrStein}, even though the scalar field in the models  they
considered was not in a state of thermal equilibrium with other
particles. It took several years until it finally became clear
that the idea of chaotic initial conditions is most general, and
it is much easier to construct a consistent cosmological theory
without making unnecessary assumptions about thermal equilibrium
and high temperature phase transitions in the early universe.

The issue of thermal initial conditions played the central role in the  long debate  about new inflation versus chaotic inflation in the 80's \cite{book}, but now  the debate is over: no realistic versions of new inflation based on the theory of thermal phase transitions and supercooling have been proposed so far. As a result, the corresponding terminology gradually changed.  Chaotic inflation, as defined in  \cite{Chaot}, occurs in all
models with sufficiently flat potentials, including the potentials
with a flat maximum, originally used in new inflation \cite{Linde:cd}.  Now the versions of  inflationary scenario with such potentials for simplicity are often
called `new inflation', even though inflation begins there not as
in the original new inflation scenario, but as in the chaotic inflation
scenario. A new twist in terminology was suggested very recently, when this version of chaotic inflation was called `hilltop inflation' \cite{Boubekeur:2005zm}.

\section{Hybrid inflation}
In the previous section we considered the simplest chaotic
inflation theory based on  the theory of a single scalar field
$\phi$. The models of chaotic inflation based on the theory of two
scalar fields may have some qualitatively new features. One of the
most interesting models of this kind is the hybrid inflation
scenario \cite{Hybrid}.  The simplest version of this scenario is
based on chaotic inflation in the theory of two scalar fields with
the effective potential
\begin{equation}\label{hybrid}
V(\sigma,\phi) =  {1\over 4\lambda}(M^2-\lambda\sigma^2)^2 +
{m^2\over 2}\phi^2 + {g^2\over 2}\phi^2\sigma^2\ .
\end{equation}
The effective mass squared of the field $\sigma$ is equal to $-M^2
+ g^2\phi^2$.  Therefore for $\phi > \phi_c = M/g$ the only
minimum of the effective potential $V(\sigma,\phi)$ is at $\sigma
= 0$. The curvature of the effective potential in the
$\sigma$-direction is much greater than in the $\phi$-direction.
Thus  at the first stages of expansion of the universe the field
$\sigma$ rolled down to $\sigma = 0$, whereas the field $\phi$
could remain large for a much longer time.

At the moment when the inflaton field $\phi$ becomes smaller than
$\phi_c = M/g$,  the phase transition with the symmetry breaking
occurs. The fields rapidly fall to the absolute minimum of the potential at $\phi = 0, \sigma^{2 } = M^{2}/\lambda$. If $m^2 \phi_c^2 = m^2M^2/g^2 \ll M^4/\lambda$, the Hubble
constant at the time of the phase transition is given by $H^2 =
{M^4 \over 12 \lambda}$ (in units $M_p = 1$). If  
$M^2 \gg {\lambda m^2\over g^2}$ and  $m^2 \ll H^2$, then the
universe at $\phi > \phi_c$ undergoes a stage of inflation, which
abruptly ends at $\phi = \phi_c$.

One of the advantages of this scenario is the possibility to
obtain small density  perturbations even if coupling constants are
large, $\lambda, g  = O(1)$, and if the inflaton field $\phi$ is much smaller than $M_{p}$. The last condition is absolutely unnecessary in the usual theory of  scalar fields coupled to gravity,  but it may be important if the scalar field has a certain geometric meaning, which is often the case in supergravity and string theory. This
makes hybrid inflation an attractive playground for those who
wants to achieve inflation in supergravity and string theory. We
will return to this question later.

\section{Quantum fluctuations and density perturbations
\label{Perturb}}

The vacuum structure in the  exponentially expanding universe is
much more complicated than in ordinary Minkowski space.
 The wavelengths of all vacuum
fluctuations of the scalar field $\phi$ grow exponentially during
inflation. When the wavelength of any particular fluctuation
becomes greater than $H^{-1}$, this fluctuation stops oscillating,
and its amplitude freezes at some nonzero value $\delta\phi (x)$
because of the large friction term $3H\dot{\phi}$ in the equation
of motion of the field $\phi$\@. The amplitude of this fluctuation
then remains almost unchanged for a very long time, whereas its
wavelength grows exponentially. Therefore, the appearance of such
a frozen fluctuation is equivalent to the appearance of a
classical field $\delta\phi (x)$ that does not vanish after
averaging over macroscopic intervals of space and time.

Because the vacuum contains fluctuations of all wavelengths,
inflation leads to the creation of more and more new perturbations
of the classical field with wavelengths greater than $H^{-1}$\@.
The average amplitude of such perturbations generated during a
typical time interval $H^{-1}$ is given by
\cite{Vilenkin:wt,Linde:uu}
\begin{equation}\label{E23}
|\delta\phi(x)| \approx \frac{H}{2\pi}\ .
\end{equation}

These fluctuations lead to density perturbations that later
produce galaxies. The theory of this effect  is very complicated
\cite{Mukh,Hawk}, and it was fully understood only in the second
part of the 80's \cite{Mukh2}. The main idea can be described as
follows:

Fluctuations of the field $\phi$ lead to a local delay of the time
of the end of inflation,  $\delta t = {\delta\phi\over \dot\phi}
\sim {H\over 2\pi \dot \phi}$. Once the usual post-inflationary
stage begins, the density of the universe starts to decrease as
$\rho = 3 H^2$, where $H \sim t^{-1}$. Therefore a local delay of
expansion leads to a local density increase $\delta_H$ such that
$\delta_H \sim \delta\rho/\rho \sim  {\delta t/t}$. Combining
these estimates together yields the famous result
\cite{Mukh,Hawk,Mukh2}
\begin{equation}\label{E24}
\delta_H \sim \frac{\delta\rho}{\rho} \sim {H^2\over 2\pi\dot\phi}
\ .
\end{equation}
This derivation is oversimplified; it does not tell, in
particular, whether $H$ should be calculated during inflation or
after it. This issue was not very important for new inflation
where $H$ was nearly constant, but it is of crucial importance for
chaotic inflation.

The result of a more detailed investigation \cite{Mukh2} shows
that $H$ and $\dot\phi$ should be  calculated during inflation, at
different times for perturbations with different momenta $k$. For
each of these perturbations the value of $H$ should be taken at
the time when the wavelength of the perturbation  becomes of the
order of $H^{-1}$. However, the field $\phi$ during inflation
changes very slowly, so the quantity ${H^2\over 2\pi\dot\phi}$
remains almost constant over exponentially large range of
wavelengths. This means that the spectrum of perturbations of
metric is flat.

A detailed calculation in our simplest chaotic inflation model the
amplitude of perturbations gives
\begin{equation}\label{E26}
\delta_H \sim   {m \phi^2\over 5\pi \sqrt 6} \ .
\end{equation}
The perturbations on scale of the horizon were produced at
$\phi_H\sim 15$ \cite{book}. This, together  with COBE
normalization $\delta_H \sim 2 \times 10^{-5}$  gives $m \sim
3\times 10^{-6}$, in Planck units, which is approximately
equivalent to $7 \times 10^{12}$ GeV. An exact value of $m$ depends on
$\phi_H$, which in its turn depends slightly on the subsequent
thermal history of the universe.

When the fluctuations of the scalar field $\phi$ are first produced (frozen), their wavelength is given by $H(\phi)^{-1}$. At the end of inflation, the wavelength grows by the factor of $e^{\phi^2/4}$, see Eq. (\ref{E04aa}). In other words,  the logarithm of the wavelength $l$ of the perturbations of metric is proportional to the value of $\phi^2$ at the moment when these perturbations were produced. As a result, according to  (\ref{E26}), the amplitude of perturbations of metric depends on the wavelength  logarithmically: $\delta_H \sim   {m\, \ln l} $. A similar logarithmic dependence (with different powers of the logarithm) appears in other versions of chaotic inflation with $V \sim \phi^{n}$ and in the simplest versions of new inflation.

Since the
observations provide us with  information about a rather limited
range of $l$, it is often possible to parametrize the scale dependence
of density perturbations by a simple power law, $\delta_H \sim
l^{(1-n_{s})/2}$. An exactly flat spectrum, called Harrison-Zeldovich spectrum, would correspond to $n_{s} = 1$.

Flatness of the spectrum of density perturbations,  together with
flatness of the universe  ($\Omega = 1$),  constitute the two most
robust predictions of inflationary cosmology. It is possible to
construct models where $\delta_H$ changes in a very peculiar way,
and it is also possible to construct theories where $\Omega \not =
1$, but it is difficult to do so.

However, there is a subtle but important difference between the prediction of flatness of the universe and the flatness of the spectrum of perturbations of metric. It is very difficult (though possible)  to construct an inflationary model deviating from the simple prediction $\Omega = 1$. Meanwhile, the situation with the flatness of the spectrum is  opposite:   it is very difficult (though possible) to construct a model with an exactly flat spectrum of perturbations of metric. In this sense, existence of a small deviation of the spectrum of inflationary perturbations from the flat Harrison-Zeldovich spectrum (i.e. the breaking of the scale invariance of the spectrum)  represents an additional robust prediction of inflation.

The small deviation from the scale invariance discussed above is fundamentally important for understanding of the global structure of inflationary universe. Here an analogy with a different branch of physics may be rather helpful.

This year we celebrated a discovery of asymptotic freedom \cite{Gross:1973id,Politzer:1973fx}, which is based, in the simplest cases, on the logarithmic growth of the coupling constants with distance. If one ignores this logarithmic dependence, one could incorrectly say that the theory is scale-invariant.

Similarly, in the simplest models of inflation the amplitude of density perturbations logarithmically increases with distance. If one ignores this logarithmic dependence, one obtains the flat Harrison-Zeldovich spectrum.

In QCD, the logarithmic dependence leads to the asymptotic freedom on small scale and to the growth of the coupling on large scale, which results in a nonperturbative regime and quark confinement. In inflationary cosmology, a similar logarithmic dependence leads to the growth of density perturbations on large scales, which results in a nonperturbative regime, fractal structure of the universe, and eternal inflation, which we are going to discuss in the next section. In both cases, the deviation of the scale invariance appears to be profoundly important. That is why it would be very interesting to find a possible deviation of the index $n_{s}$ from 1. Hopefully, this can be achieved in  future observations; in particular, Planck satellite may be capable of measuring the deviation $n_{s}-1$ with an accuracy about 0.5\%.

For future reference, we will give here a list of equations which are often used for comparison of predictions of inflationary theories with observations in the slow roll approximation.

The amplitude of scalar perturbations of metric can be characterized either by $\delta_{H}$, or by a closely related quantity $\Delta_{\cal R}$ \cite{LL}. Similarly, the amplitude of tensor perturbations is given by $\Delta_h$.  Following 
\cite{LL,WMAP,Tegmark,Seljak}, one can represent these quantities as 
\begin{eqnarray}
 \label{eq:P_R} \Delta^2_{\cal R}(k)&=& \Delta^2_{\cal R}(k_0)
  \left(\frac{k}{k_0}\right)^{n_s-1}, \\ \label{eq:P_h} \Delta^2_h(k)&=& \Delta^2_h(k_0)
  \left(\frac{k}{k_0}\right)^{n_t},
\end{eqnarray}
where $\Delta^2(k_0)$ is a normalization constant, and $k_0$ is a normalization point. Here we ignored running of the indexes $n_{s}$ and $n_{t}$ since there is no observational evidence that it is significant.

One can also introduce the tensor/scalar ratio
$r$, the relative amplitude  of the tensor to  scalar modes,
\begin{equation}
 \label{eq:rdef} r \equiv \frac{\Delta^2_h(k_0)}{\Delta^2_{\cal R}(k_0)}.
\end{equation}

There
are three slow-roll
parameters \cite{LL}
\begin{eqnarray}
 \label{eq:eps} \epsilon =
  \frac{1}{2}\left(\frac{V'}{V}\right)^2, ~~
  \eta  =  \frac{V''}{V}, ~~
  \label{eq:xi} \xi  =
\frac{V'V'''}{V^2},
\end{eqnarray}
where prime denotes derivatives with respect to the field
$\phi$. All parameters must be
smaller than one for the slow-roll approximation to be valid. 

 A standard slow roll analysis gives observable
quantities in terms of the slow roll parameters to first order as
\begin{eqnarray}
 \label{eq:A} &&\Delta^2_{\cal R}   =  \frac{V}{24\pi^2\epsilon} =  \frac{V^{3}}{12\pi^2(V')^{2}},\\
    \label{eq:n_s} &&n_s-1 =
  -6\epsilon + 2\eta , \\
  \label{eq:r} &&r   =  16 \epsilon, \\
  \label{eq:n_t} &&n_t   =  -2\epsilon = -\frac{r}{8}.
\end{eqnarray}
 The equation
$n_t=-r/8$ is known as the consistency relation for single-field
inflation models; it becomes an inequality for multi-field
inflation models. If $V$ during inflation is sufficiently large, as in the simplest models of chaotic inflation, one may have a chance to find the tensor contribution to the CMB anisotropy. The possibility to determine $n_{t}$ is less certain. 
It may be rather difficult, though maybe not impossible \cite{Sigurdson:2005cp}, to find the tensor contribution in new inflation and in hybrid inflation.
The most important information which can be obtained now from the cosmological observations at present is related to Eqs. (\ref{eq:A}) and (\ref{eq:n_s}).

Following notational conventions in \cite{WMAP}, we use
$A(k_0)$ for the scalar power spectrum amplitude, where $A(k_0)$ and
$\Delta^2_{\cal R}(k_0)$ are related through
\begin{eqnarray}
 \label{eq:Adef} \Delta^2_{\cal R}(k_0) \simeq 3\times10^{-9}  A(k_0).
\end{eqnarray}
The parameter $A$ is often normalized at $k_{0} \sim 0.05$/Mpc; its observational value is about 0.8 \cite{WMAP,Tegmark}. This leads to the observational constraint on $V(\phi)$ following from the normalization of the spectrum of the large-scale density perturbations:
\begin{eqnarray}
 \label{eq:V} {V^{{3/2}}\over V'} \simeq 5\times10^{-4} \ .
\end{eqnarray}
Here $V(\phi)$ should be evaluated for the value of the field $\phi$ which is determined by the condition that the perturbations produced at the moment when the field was equal $\phi$  evolve into the present time perturbations with momentum $k_{0} \sim 0.05$/Mpc. In the first approximation, one can find the corresponding moment by assuming that it happened 60 e-foldings before the end of inflation. The number of e-foldings can be calculated in the slow roll approximation using the relation
\begin{eqnarray}
 \label{eq:N} {N} \simeq \int_{\phi_{\rm end}}^{\phi}{V\over V'} d\phi \ .
\end{eqnarray}
Finally, recent observational data imply \cite{Seljak} that 
\begin{eqnarray}
 \label{eq:ns} n_s =1 
  -{3}\left(\frac{V'}{V}\right)^2 + 2\frac{V''}{V} = 0.98 \pm  0.03 \ .
\end{eqnarray}
These relations are very useful for comparing inflationary models with observations.
In particular, the simplest versions of chaotic and new inflation predict $n_{s} < 1$, whereas in hybrid inflation one may have either $n_{s} < 1$ or $n_{s}>  1$, depending on the model.

Here we concentrated on the simplest and most general mechanism of production of adiabatic perturbations of metric in inflationary cosmology. However, one should keep in mind that quantum fluctuations produced during inflation may also lead to isocurvature fluctuations \cite{Linde:1985yf}, which may later convert into adiabatic perturbations (the so-called curvaton mechanism \cite{Lyth:2001nq,Moroi:2001ct,Mollerach:1989hu,Linde:1996gt}). One may also produce adiabatic perturbations by perturbing effective coupling constants, which modulates  the process of reheating \cite{Dvali:2003em,Kofman:2003nx}.

\section{Eternal inflation}

A significant step in the development of inflationary theory was
the discovery of the process of self-reproduction of inflationary
universe. This process was known to exist in old inflationary
theory \cite{Guth} and in the new one \cite{StLin,Vilenkin:xq}, but its
significance was fully realized only after the discovery of the
regime of eternal inflation in the simplest versions of the
chaotic inflation scenario \cite{Eternal,LLM}. It appears that in
many models large quantum fluctuations produced during inflation
which may locally increase the value of the energy density in some
parts of the universe. These regions expand at a greater rate than
their parent domains, and quantum fluctuations inside them lead to
production of new inflationary domains which expand even faster.
This  leads to an eternal process of self-reproduction of the
universe.

To understand the mechanism of self-reproduction one should
remember that the processes separated by distances $l$ greater
than $H^{-1}$ proceed independently of one another. This is so
because during exponential expansion the distance between any two
objects separated by more than $H^{-1}$ is growing with a speed
exceeding the speed of light. As a result, an observer in the
inflationary universe can see only the processes occurring inside
the horizon of the radius  $H^{-1}$. An important consequence of
this general result is that the process of inflation in any
spatial domain of radius $H^{-1}$ occurs independently of any
events outside it. In this sense any inflationary domain of
initial radius exceeding $H^{-1}$ can be considered as a separate
mini-universe.

To investigate the behavior of such a mini-universe, with an
account taken of quantum fluctuations, let us consider an
inflationary domain of initial radius $H^{-1}$ containing
sufficiently homogeneous field with initial value $\phi \gg M_p$.
Equation (\ref{E04}) implies that during a typical time interval
$\Delta t=H^{-1}$ the field inside this domain will be reduced by
$\Delta\phi = \frac{2}{\phi}$. By comparison this expression with
$|\delta\phi(x)| \approx \frac{H}{2\pi} =  {m\phi\over 2\pi\sqrt
6}$ one can easily see that if $\phi$ is much less than $\phi^*
\sim {5\over  \sqrt{ m}} $,
 then the decrease of the field $\phi$
due to its classical motion is much greater than the average
amplitude of the quantum fluctuations $\delta\phi$ generated
during the same time. But for   $\phi \gg \phi^*$ one has
$\delta\phi (x) \gg \Delta\phi$. Because the typical wavelength of
the fluctuations $\delta\phi (x)$ generated during the time is
$H^{-1}$, the whole domain after $\Delta t = H^{-1}$ effectively
becomes divided into $e^3 \sim 20$ separate domains
(mini-universes) of radius $H^{-1}$, each containing almost
homogeneous field $\phi - \Delta\phi+\delta\phi$.   In almost a
half of these domains the field $\phi$ grows by
$|\delta\phi(x)|-\Delta\phi \approx |\delta\phi (x)| = H/2\pi$,
rather than decreases. This means that the total volume of the
universe containing {\it growing} field $\phi$ increases 10 times.
During the next time interval $\Delta t = H^{-1}$ this process
repeats. Thus, after the two time  intervals $H^{-1}$ the total
volume of the universe containing the growing scalar field
increases 100 times, etc. The universe enters eternal process of
self-reproduction.

The existence of this process implies that the  universe will never disappear as a whole. Some of its parts may collapse, the life in our part of the universe may perish, but there always will be some other parts of the universe where life will appear again and again, in all of its possible forms.

One should be careful, however, with the interpretation of these results. There is still an ongoing debate of whether eternal inflation is eternal only in the future or also in the past. In order to understand what is going on, let us consider any particular time-like geodesic line at the stage of inflation. One can show that for any given observer following this geodesic, the duration $t_{i}$ of the stage of inflation on this geodesic will be finite. One the other hand, eternal inflation implies that if one takes all such geodesics and calculate the time $t_{i}$ for each of them, then there will be no upper bound for $t_{i}$, i.e. for each time $T$ there will be such geodesic which experience inflation for the time $t_{i} >T$.  Even though the relative number of long geodesics can be very small, exponential expansion of space surrounding them will lead to an eternal exponential growth of the total volume of inflationary parts of the universe.

Similarly, if one concentrates on any particular geodesic in the past time direction, one can prove that it has finite length \cite{Borde:2001nh}, i.e. inflation in  any particular point of the universe should have a beginning at some time $\tau_{i}$. However, there is no reason to expect that there is an upper bound for all $\tau_{i}$ on all geodesics. If this upper bound does not exist, then eternal inflation is eternal not only in the future but also in the past.

In other words, there was a  beginning for each part of the universe, and there will be an end for inflation at any particular point. But there will be no end for the evolution of the universe {\it as a whole} in the eternal inflation scenario, and at present we do not have any reason to believe that there was a single beginning of the evolution of the whole universe at some moment $t = 0$, which was traditionally associated with the Big Bang.

To illustrate the process of eternal inflation, we present here the results of computer simulations of evolution of a system of two scalar fields during inflation. The field $\phi$ is the inflaton field driving inflation; it is shown by the height of the distribution of the field $\phi(x,y)$ in a two-dimensional slice of the universe. The second field, $\Phi$, determines the type of spontaneous symmetry breaking which may occur in the theory. We paint the surface black if this field is in a state corresponding to one of the two minima of its effective potential;  we paint it white if it is in the second minimum corresponding to a different type of symmetry breaking, and therefore to a different set of laws of low-energy physics.

\begin{figure}

\centering\leavevmode\epsfysize=13cm \epsfbox{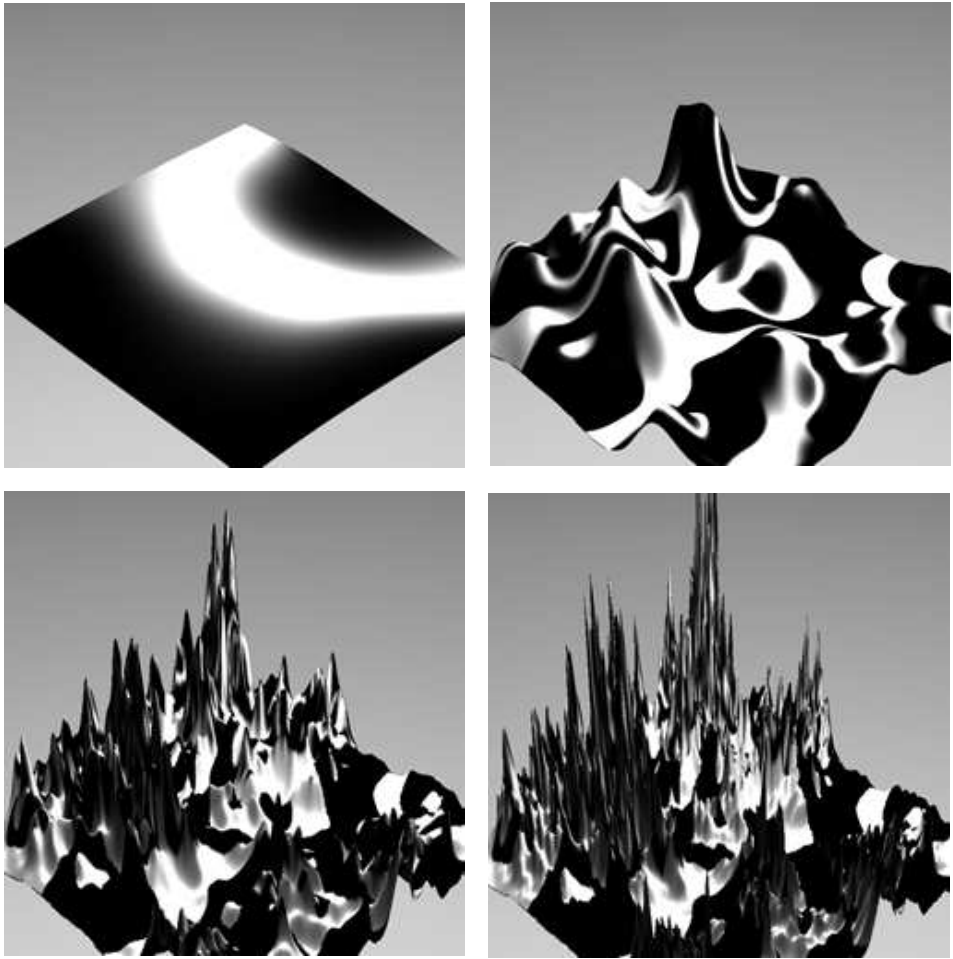}

\

\caption{Evolution of scalar fields  $\phi$ and $\Phi$ during the process of self-reproduction of the universe.   The height of the distribution shows the value of the field $\phi$ which drives inflation. The surface is painted black in those parts of the universe where the scalar field $\Phi$ is in the first minimum of its effective potential, and  white where it is in the second minimum. Laws of low-energy physics are different in the regions of different color. The peaks of the ``mountains'' correspond  to places where quantum fluctuations bring the scalar fields back to the Planck density. Each of such places in a certain sense can be considered as a beginning of a new Big Bang. }
\label{fig:Fig0}
\end{figure}

In the beginning of the process the whole inflationary domain is black, and the distribution of both fields is very homogeneous. Then the domain became exponentially large (but it has the same size in comoving coordinates, as shown in Fig. \ref{fig:Fig0}).  Each peak of the mountains corresponds to nearly Planckian density and can be interpreted as a beginning of a new ``Big Bang.'' The laws of physics are rapidly changing there, but they become fixed in the parts of the universe where the field $\phi$ becomes small. These parts correspond to valleys in Fig. \ref{fig:Fig0}. Thus quantum fluctuations of the scalar fields divide the universe into exponentially large domains with different laws of low-energy physics, and with different values of energy density. 

If this scenario is correct, then physics alone cannot provide a 
complete explanation for all properties of our part of the universe.   
The same physical theory may yield large parts of the universe that 
have diverse properties.  According to this scenario, we find 
ourselves inside a four-dimensional domain with our kind of 
physical laws not because domains with different dimensionality 
and with alternate properties are impossible or improbable, but 
simply because our kind of life cannot exist in other domains.

 The fact that inflation may happen in a different manner in different parts of the universe was recognized at the very early stages of development of inflationary theory, which allowed us to justify the use of anthropic principle in inflationary cosmology \cite{Linde:1984je}. Eternal inflation makes it possible to go even further: It implies that even if the universe started in a single domain with well defined initial conditions, the process of eternal inflation will divide it into infinitely many exponentially large domains that have
different laws of low-energy physics \cite{Eternal,LLM}. Among all of these domains, we can live and make observations only in those that are compatible with our existence.

Eternal inflation scenario was extensively studied during the last 20 years. I should mention, in particular, the discovery of the topological eternal inflation \cite{TopInf}, calculation of the fractal dimension of the universe \cite{Aryal:1987vn,LLM}, and development of various methods describing statistical/probabilistic aspects of the  self-reproducing universe, see  \cite{LLM,Garcia-Bellido:1993wn,Vilenkin:1995yd,Tegmark:2004qd} and references therein. This scenario may have  especially interesting implications in the context
of string theory, which allows exponentially large number of
different de Sitter  vacuum states \cite{book,landscape,BP,Douglas}, see
Sect. \ref{land}.

\section{Creation of matter after inflation: reheating and preheating}

The theory of reheating of the universe after inflation is the most
important application of the quantum theory of particle creation, since
almost all matter constituting the universe  was created during this
process.

At the stage of inflation all energy is concentrated in a classical
slowly moving inflaton field $\phi$. Soon after the end of inflation this
field begins to oscillate near the minimum of its effective potential.
Eventually it produces many elementary particles, they interact  with
each other and come to a state of thermal equilibrium with some
temperature $T_r$.

Early discussions of reheating of the universe after inflation  \cite{oldtheory} were based on the idea that the homogeneous inflaton field can be represented as a collection of the particles of the field $\phi$. Each of these particles decayed independently. This process can be studied by the usual perturbative approach to particle decay. Typically, it takes thousands of oscillations of the inflaton field until it decays into usual elementary particles by this mechanism.  More recently, however, it was discovered that coherent field effects such as parametric resonance can lead to the decay of the homogeneous field much faster than would have been predicted by perturbative methods, within few dozen oscillations \cite{KLS}. These coherent effects produce high energy, nonthermal fluctuations that could have
significance for understanding developments in the early universe, such as baryogenesis.  This early stage of rapid nonperturbative decay  was called `preheating.' 
In \cite{tach} it was found that another effect known as tachyonic preheating can lead to even faster decay than parametric
resonance. This effect occurs whenever the homogeneous field rolls down a tachyonic ($V''<0$) region of its
potential. When that occurs, a tachyonic, or spinodal instability leads to exponentially rapid growth of all long wavelength modes with 
$k^2<|V''|$. This growth can often drain all of the energy from the homogeneous field within a single oscillation.

We are now in a position to classify the dominant mechanisms by which the homogeneous inflaton field decays in different classes of
inflationary models. Even though all of these models, strictly speaking,  belong to the general class of chaotic inflation (none of them is based on the theory of thermal initial conditions), one can  break them into three classes: small field, or new inflation models \cite{New}, large field, or chaotic inflation models of the type of the model $m^2\phi^2/2$ \cite{Chaot}, and multi-field, or hybrid models \cite{Hybrid}. This classification is incomplete, but still rather helpful.

In the simplest versions of chaotic inflation, the stage of preheating is generally dominated by
parametric resonance, although there are parameter ranges where this
can not occur \cite{KLS}.    In \cite{tach} it was shown that tachyonic preheating
dominates the preheating phase in hybrid models of inflation. New inflation in this respect occupies an intermediate position between chaotic inflation and hybrid inflation:  If spontaneous symmetry breaking in this scenario is very large, reheating occurs due to parametric resonance and perturbative decay. However, for the models with spontaneous symmetry breaking at or below the GUT scale, $\phi \ll 10^{{-2}} M_p$, preheating occurs due to a combination of tachyonic preheating and parametric resonance. The resulting effect is very strong, so that the homogeneous mode of the inflaton field typically decays within few oscillations \cite{Desroche:2005yt}. 

A detailed investigation of preheating usually requires lattice simulations, which can be achieved following \cite{latticeold,latticeeasy}.  Note that preheating is not the last stage of reheating; it is followed by a period of turbulence \cite{thermalization}, by a much slower perturbative decay described by the methods developed in \cite{oldtheory}, and by eventual thermalization.

\section{Inflation and observations}

Inflation is not just an interesting theory that can resolve many
difficult problems of the standard  Big Bang cosmology. This
theory made several   predictions which can be tested by
cosmological observations. Here are the most important
predictions:

1) The universe must be flat. In most models $\Omega_{total} = 1
\pm 10^{-4}$.

2) Perturbations of metric produced during inflation are
adiabatic.

3) Inflationary perturbations have nearly flat spectrum.  In most
inflationary models the spectral index $n_{s} = 1 \pm 0.2$ ($n_{s}=1$
means totally flat).

4) The spectrum of inflationary perturbations should be slightly non-flat. (It is very difficult to construct a model with  $n_{s}   =1$.) 

5) These perturbations are gaussian.

6) Perturbations of metric could be scalar, vector or tensor.
Inflation mostly produces scalar  perturbations, but it also
produces tensor perturbations with nearly flat spectrum, and it
does {\it not} produce vector perturbations. There are certain
relations between the properties of  scalar and tensor
perturbations produced by inflation.

7) Inflationary perturbations produce specific peaks in the
spectrum of CMB radiation. (For a simple pedagogical
interpretation of this effect see  e.g. \cite{Dodelson:2003ip}; a
detailed theoretical description can be found in
\cite{Mukhanov:2003xr}.)

It is possible to violate each of these predictions if one makes
this theory sufficiently complicated. For example, it is possible
to produce vector perturbations of metric in the models where
cosmic strings are produced at the end of inflation, which is the
case in some versions of hybrid inflation. It is possible to have
an open or closed inflationary universe, or even a small periodic
inflationary universe, it is possible to have models with
nongaussian isocurvature fluctuations with a non-flat spectrum.
However, it is very difficult to do so, and most of the
inflationary models obey the simple rules given above.

\begin{figure}

\centering\leavevmode\epsfysize=10cm \epsfbox{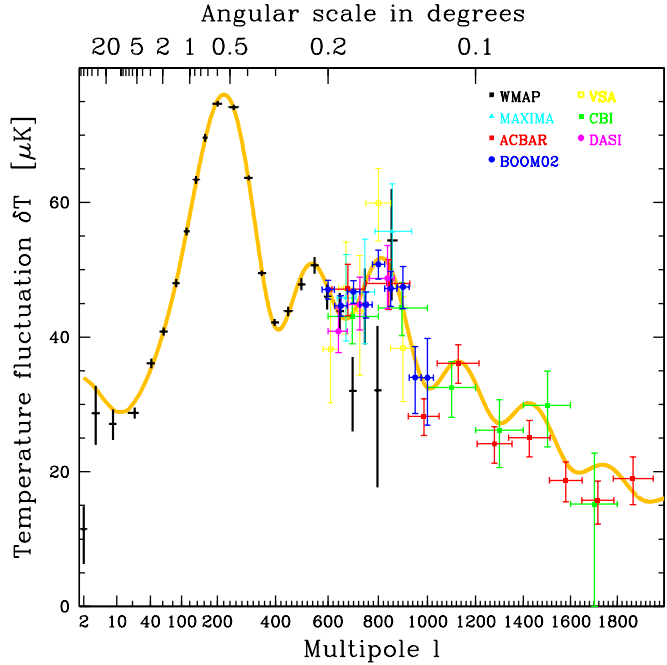}

\caption{CMB data versus the predictions of one of the simplest inflationary models (thick yellow line), according to \cite{Tegmark}.}
\label{cmb}
\end{figure}

It is not easy to test all of these predictions. The major
breakthrough in this direction was achieved  due to the recent
measurements of the CMB anisotropy. The latest results based on
the WMAP experiment, in combination with the Sloan Digital Sky
Survey, are consistent with predictions of the simplest
inflationary models with adiabatic gaussian perturbations, with
$\Omega = 1.01 \pm 0.02$, and $n_{s} = 0.98 \pm
0.03$~\cite{WMAP,Tegmark,Seljak}, see Fig. \ref{cmb}.

There are still some question marks to be examined, such as an
unexpectedly small anisotropy of CMB at  large angles \cite{WMAP} and possible correlations between low multipoles.
However,  the observational status and interpretation of these effects is rather uncertain \cite{Efstathiou:2003wr}, and it is quite significant that all proposed
explanations of these observations are based on inflationary cosmology,
see e.g. \cite{Contaldi}.

\section{Alternatives to inflation?}\label{alt}

Inflationary scenario is very versatile, and now, after 20 years
of persistent attempts of many physicists to propose an
alternative to inflation, we still do not know any other  way to
construct a consistent cosmological theory. Indeed, in order to
compete with inflation a new theory should make similar
predictions and should offer an alternative solution to many
difficult cosmological problems. Let us look at these problems
before starting a discussion.

1) Homogeneity problem. Before even starting investigation of
density perturbations and structure  formation, one should explain
why the universe is nearly homogeneous on the horizon scale.

2) Isotropy problem. We need to understand why all directions in
the universe are similar to each  other, why there is no overall
rotation of the universe. etc.

3) Horizon problem. This one is closely related to the homogeneity
problem. If different parts of  the universe have not been in a
causal contact when the universe was born, why do they look so
similar?

4) Flatness problem. Why $\Omega \approx 1$? Why parallel lines do
not intersect?

5) Total entropy problem. The total entropy of the observable part
of the universe is  greater than $10^{87}$. Where did this huge
number come from? Note that the lifetime of a closed universe
filled with hot gas with total entropy $S$  is $S^{2/3}\times
10^{-43}$ seconds \cite{book}. Thus $S$ must be huge. Why?

6) Total mass problem. The total mass of the observable part of
the universe has mass  $\sim 10^{60} M_p$.  Note also that the
lifetime of a closed universe filled with nonrelativistic
particles of total mass $M$ is ${M\over M_P} \times 10^{-43}$
seconds. Thus $M$ must be huge. But why?

7) Structure formation problem. If we manage to explain the
homogeneity of the universe, how can  we explain the origin of
inhomogeneities required for the large scale structure formation?

8) Monopole problem, gravitino problem, etc.

This list is very long. That is why it was not easy to propose any
alternative to inflation even  before we learned that $\Omega
\approx 1$, $n_{s}\approx 1$, and that the perturbations responsible
for galaxy formation are mostly adiabatic, in agreement with the
predictions of the simplest inflationary models.

Despite this difficulty, there was always a tendency to announce
that we have  eventually found a good alternative to inflation.
This was the ideology of the models of structure formation due to
topological defects, or textures, which were often described as
competitors to inflation, see e.g. \cite{SperTur}. However, it was
clear from the very beginning that these theories at best could
solve only one problem (structure formation) out of 8 problems
mentioned above. Therefore the true question was not whether one
can replace inflation by the theory of cosmic strings/textures,
but whether inflation with cosmic strings/textures is better than
inflation without cosmic strings/textures. Recent observational
data favor the simplest version of inflationary theory, without
topological defects, or with an extremely small (few percent)
admixture of the effects due to cosmic strings.

A similar situation emerged with the introduction of the
ekpyrotic  scenario  \cite{KOST}. In the original version of this
theory it was claimed that this scenario can solve all
cosmological problems without using the stage of inflation, i.e.
without a prolonged stage of an accelerated expansion of the
universe, which was called in \cite{KOST} ``superluminal
expansion.'' However, this original idea did not work
\cite{KKL,KKLTS}, and the idea to avoid ``superluminal expansion''
was abandoned by the authors of \cite{KOST}. A more recent version
of this scenario, the cyclic scenario \cite{cyclic}, uses an
infinite number of periods of ``superluminal expansion'', i.e.
inflation, in order to solve the major cosmological problems. In this sense, the cyclic scenario is not a true alternative to inflationary scenario, but its rather peculiar version. The main difference between the usual inflation and the cyclic
inflation, just as in the case of topological defects and
textures, is the mechanism of generation of density perturbations.
However, since the theory of density perturbations in cyclic
inflation requires a solution of the cosmological singularity
problem \cite{Liu:2002ft,Horowitz:2002mw}, it is difficult to say
anything definite about it. The latest attempts to do so (despite our inability to address the singularity problem) indicate that the spectrum of metric perturbations produced in the cyclic scenario is incompatible with observations \cite{Creminelli:2004jg,Bozza:2005wn}.

Thus at the moment it is hard to see any real alternative to
inflationary cosmology; instead of a competition between inflation
and other ideas, we witness a competition between many different
models of inflationary theory.

This competition goes in several different directions. First of
all, we must try to implement inflation in realistic theories of
fundamental interactions. But what do we mean by `realistic?'
Here we have an interesting and even somewhat paradoxical situation.
In the absence of a direct confirmation of M/string theory and
supergravity by high energy physics experiments (which may change
when we start receiving data from the LHC), the definition of what
is realistic becomes increasingly dependent on cosmology and the
results of the cosmological observations. In particular, one may
argue that those versions of the theory of all fundamental
interactions that cannot describe inflation and the present stage
of acceleration of the universe are disfavored by observations.

On the other hand, not every theory which can lead to inflation
does it in an equally good way. Many inflationary models have been
already ruled out be observations. This happened long ago with
such models as extended inflation \cite{extended} and the simplest
versions of ``natural inflation'' \cite{natural}. Recent data from
WMAP and SDSS almost ruled out a particular version of chaotic
inflation with $V(\phi) \sim \phi^4$ \cite{WMAP,Tegmark}.

However, observations test only the last stages of inflation. In
particular, they  do not say anything about the properties of the
inflaton potential at $V(\phi) \gtrsim 10^{-10} M_p^4$. Thus there
may exist many different models which describe all observational
data equally well. In order to compare such models, one should not
only compare their predictions with the results of the
cosmological observations, but also carefully examine whether these models are internally consistent and whether it is possible to solve the problem of initial conditions for inflation in these models. One should also try to find out the best way to implement inflationary scenario in the context of realistic models of all fundamental interactions, including the models based on supergravity and string theory.

\section{Shift symmetry and chaotic inflation in supergravity}

Most of the existing inflationary models are based on the idea of
chaotic initial conditions, which is the trademark of the chaotic
inflation scenario. In the simplest versions of chaotic inflation
scenario with the potentials $V \sim \phi^n$, the process of
inflation occurs at $\phi>1$, in Planck units. Meanwhile, there
are many other models where inflation may occur at $\phi \ll 1$.

There are several reasons why this difference may be important.
First of all, some authors argue that  the generic expression for
the effective potential can be cast in the form
\begin{equation}\label{LythRiotto}
V(\phi) = V_0 +\alpha \phi+ {m^2\over 2} \phi^2 +{\beta\over 3}
\phi^3+ {\lambda\over 4} \phi^4 + \sum_n \lambda_n
{\phi^{4+n}\over {M_p}^n}\, ,
\end{equation}
and then they assume that generically $\lambda_n = O(1)$, see e.g.
Eq. (128) in \cite{LythRiotto}.  If this assumption were correct,
one would have little control over the behavior of $V(\phi)$ at
$\phi > M_p$.

Here we have written $M_p$ explicitly, to expose the implicit
assumption made in \cite{LythRiotto}.  Why do we write $M_p$ in
the denominator, instead of $1000 M_p$? An intuitive reason is
that quantum gravity is non-renormalizable, so one should
introduce a cut-off at momenta $k \sim M_p$. This is a reasonable
assumption, but it does not imply validity of Eq.
(\ref{LythRiotto}). Indeed, the constant part of the scalar field
appears in the gravitational diagrams not directly, but only via
its effective potential $V(\phi)$ and the masses of particles
 interacting with the scalar field $\phi$. As a result,
the terms induced by quantum gravity effects are suppressed not by
factors ${\phi^n \over {M_p}^n}$, but by factors  $V\over {M_p}^4$
and $m^2(\phi)\over {M_p}^2$ \cite{book}. Consequently, quantum
gravity corrections to $V(\phi)$ become large not at $\phi
> M_p$, as one could infer from (\ref{LythRiotto}), but only at
super-Planckian energy density, or for super-Planckian masses.
This justifies our use of the simplest chaotic inflation models.

The simplest way to understand this argument is to consider again the
case where the potential of the field $\phi$ is a constant,
$V=V_0$. Then the theory has a {\it shift symmetry}, $\phi \to
\phi +c$. This symmetry is not broken by perturbative quantum
gravity corrections, so no such terms as $\sum_n \lambda_n
{\phi^{4+n}\over {M_p}^n}$ are generated. This symmetry may be
broken by nonperturbative quantum gravity effects (wormholes?
virtual black holes?), but such effects, even if they exist, can
be made exponentially small \cite{Kallosh:1995hi}.

The idea of shift symmetry appears to be very fruitful in
application to inflation; we will return to it many times in this
paper. However, in some cases the scalar field $\phi$ itself may
have physical (geometric) meaning,  which may constrain the
possible values of the fields during inflation. The most important
example is given by $N = 1$ supergravity.

The effective potential of the complex scalar field $\Phi$ in
supergravity is given by the well-known  expression (in units $M_p
= 1$):
\begin{equation}\label{superpot}
V = e^{K} \left[K_{\Phi\bar\Phi}^{-1}\, |D_\Phi W|^2
-3|W|^2\right].
\end{equation}
Here $W(\Phi)$ is the superpotential, $\Phi$ denotes the scalar
component of the superfield  $\Phi$; $D_\Phi W= {\partial W\over
\partial \Phi} + {\partial K\over \partial \Phi} W$. The kinetic
term of the scalar field is given by $K_{\Phi\bar\Phi}\,
\partial_\mu \Phi \partial _\mu \bar\Phi$. The standard textbook
choice of the K\"ahler potential corresponding to the canonically
normalized fields $\Phi$ and $\bar\Phi$ is $K = \Phi\bar\Phi$, so
that $K_{\Phi\bar\Phi}=1$.

This immediately reveals a problem: At $\Phi > 1$ the potential is
extremely steep.  It blows up as $e^{|\Phi|^2}$, which makes it
very difficult to realize chaotic inflation in supergravity at
$\phi \equiv \sqrt 2|\Phi| > 1$. Moreover, the problem persists
even at small $\phi$. If, for example, one considers the simplest
case when there are many other scalar fields  in the theory  and
the superpotential does not depend on the inflaton field $\phi$,
then Eq. (\ref{superpot}) implies that at $\phi \ll 1$ the
effective mass of the inflaton field is $m^2_\phi = 3H^2$. This
violates the  condition $m^2_\phi \ll H^2$ required for successful
slow-roll inflation (so-called $\eta$-problem).

The major progress in SUGRA inflation during the last decade was
achieved in the context of the models of the hybrid inflation
type, where inflation may occur at $\phi \ll 1$. Among the best
models are the F-term inflation, where different contributions to
the effective mass term $m^2_\phi$ cancel \cite{F}, and D-term
inflation \cite{D}, where the dangerous term $e^K$ does not affect
the potential in the inflaton direction. A detailed discussion of
various versions of hybrid inflation in supersymmetric theories
can be found in \cite{LythRiotto}. A recent version of this
scenario, P-term inflation, which unifies F-term and D-term
models, was proposed in \cite{pterm}.

However, hybrid inflation occurs only on a relatively small energy
scale, and many of its versions do not lead to eternal inflation.
Therefore it would be nice to obtain inflation in a context of a
more general class of supergravity models.

This goal seemed very difficult to achieve; it took almost 20
years to find a natural realization of chaotic inflation model in
supergravity. Kawasaki, Yamaguchi and Yanagida suggested to take
the K\"ahler potential
\begin{equation} K = {1\over 2}(\Phi+\bar\Phi)^2
+X\bar X \end{equation}
 of the fields $\Phi$ and $X$, with the
superpotential $m\Phi X$ \cite{jap}.

At the first glance, this K\"ahler potential may seem somewhat
unusual. However, it can be obtained from the standard K\"ahler
potential $K =  \Phi \bar\Phi  +X\bar X$  by adding terms
$\Phi^2/2+\bar\Phi^2/2$, which do not give any contribution to the
kinetic term of the scalar fields $K_{\Phi\bar\Phi}\,
\partial_\mu \Phi \partial _\mu \bar\Phi$. In other words, the new
K\"ahler potential, just as the old one, leads to canonical
kinetic terms for the fields $\Phi$ and $X$, so it is as simple
and legitimate as the standard textbook K\"ahler potential.
However, instead of the U(1) symmetry with respect to rotation of
the field $\Phi$ in the complex plane, the new K\"ahler potential
has a {\it shift symmetry}; it does not depend on the imaginary
part of the field $\Phi$. The shift symmetry is broken only by the
superpotential.

This leads to a profound  change of the potential
(\ref{superpot}): the dangerous term $e^K$ continues growing
exponentially in the direction $(\Phi +\bar\Phi)$, but it remains
constant in the direction $(\Phi - \bar \Phi )$. Decomposing the
complex field $\Phi$ into two real scalar fields, $ \Phi = {1\over
\sqrt 2} (\eta +i\phi)$, one can find the resulting potential
$V(\phi,\eta,X)$ for $\eta, |X| \ll 1$:
\begin{equation}\label{superpot1}
V = {m^2\over 2} \phi^2 (1 + \eta^2) + m^2|X|^2.
\end{equation}
This potential has a deep valley, with a minimum at $\eta = X =0$.
Therefore the fields $\eta$ and $X$ rapidly fall down towards
$\eta = X =0$, after which the potential for the field $\phi$
becomes $V = {m^2\over 2} \phi^2$. This provides  a very simple
realization of eternal chaotic inflation scenario in supergravity
\cite{jap}. This model can be extended to include theories with
different power-law potentials, or models where inflation begins
as in the simplest versions of chaotic inflation scenario, but
ends as in new or hybrid inflation, see e.g.
\cite{Yamaguchi:2001pw,Yok}.

It is amazing that for almost 20 years nothing but inertia was
keeping us from using the version of the supergravity which was
free from the famous $\eta$ problem. As we will see shortly, the
situation with inflation in string theory is very similar, and may
have a similar resolution.

\section{Towards Inflation in String Theory}
\subsection{de Sitter vacua in string theory}

For a long time, it seemed rather difficult to obtain inflation in
M/string theory. The main problem here was the stability of
compactification of internal dimensions. For example, ignoring
non-perturbative effects to be discussed below, a typical
effective potential of the effective 4d theory obtained by
compactification in string theory of type IIB can be represented
in the following form:
\begin{equation}
V(\varphi,\rho,\phi) \sim e^{\sqrt 2\varphi -\sqrt6\rho}\ \tilde
V(\phi)
\end{equation}
Here $\varphi$ and $\rho$ are canonically normalized fields
representing the dilaton field and the volume of the compactified
space; $\phi$ stays for all other fields, including the inflaton field.

If $\varphi$ and $\rho$ were constant, then the potential $\tilde
V(\phi)$ could drive inflation.  However, this does not happen
because of the steep exponent $e^{\sqrt 2\varphi -\sqrt6\rho}$,
which rapidly pushes the dilaton field $\varphi$ to $-\infty$, and
the volume modulus $\rho$ to $+\infty$. As a result, the radius of
compactification becomes infinite; instead of inflating, 4d space
decompactifies and becomes 10d.

Thus in order to describe inflation one should first learn how to
stabilize the dilaton and the volume modulus. The dilaton
stabilization was achieved in \cite{GKP}. The most difficult
problem was to stabilize the volume. The solution of this problem
was found in \cite{KKLT} (KKLT construction). It consists of two
steps.

\begin{figure}[h!]
\centering\leavevmode\epsfysize=6.5cm \epsfbox{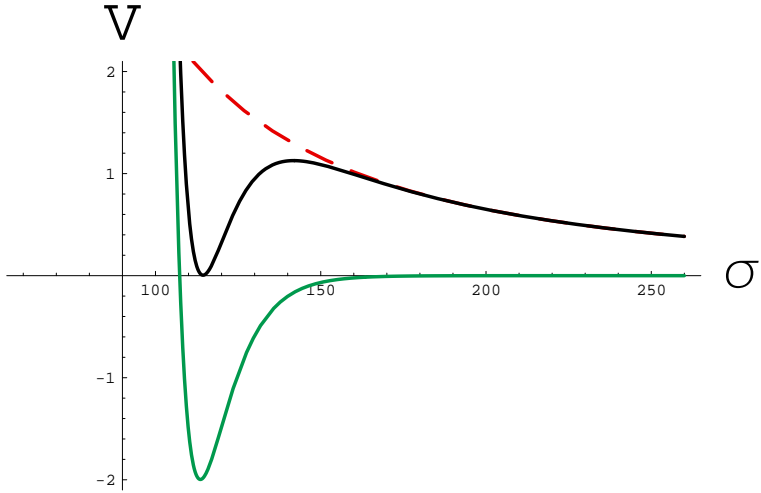} \caption[fig1]
{KKLT potential as a function of $\sigma = {\rm Re}\,\rho$. Thin green line corresponds to AdS stabilized potential for $W_0 =- 10^{{-4}}$,
$A=1$, $a =0.1$. Dashed line shows the additional term,
which appears either due to the contribution of a $\overline{D3}$ brane or of a
D7 brane. Thick black line shows the resulting potential  with a very small but positive value of $V$ in the minimum. All potentials are shown multiplied by $10^{15}$.} \label{1}
\end{figure}

First of all, due to a combination of effects related to warped
geometry of the compactified space and nonperturbative effects
calculated directly in 4d (instead of being obtained by
compactification), it was possible to obtain a supersymmetric AdS
minimum of the effective potential for $\rho$. In the original version of the KKLT scenario, it was done in the theory with the  K\"ahler potential
\begin{equation}
K = -3\log (\rho+\bar\rho)
, \end{equation} and with the nonperturbative superpotential of the form \begin{equation}\label{KKLTsp}
W=W_0+
Ae^{-a\rho}, 
\end{equation} 
with $a = 2\pi/N$.
The corresponding effective potential for the complex field $\rho =
\sigma +i\alpha$ had a minimum at finite, moderately large values of the volume modulus field $\sigma_{0}$, which  fixed the
volume modulus  in a state with a negative vacuum energy. Then
 an anti-${D3}$ brane with the positive energy $\sim
\sigma^{-2}$ was added. This addition uplifted the minimum of the potential to
the state with a positive vacuum energy, see Fig. \ref{1}.

Instead of adding an anti-${D3}$ brane, which explicitly breaks
supersymmetry, one can add  a D7 brane with fluxes. This results
in the appearance of a D-term which has a similar dependence on
$\rho$, but leads to spontaneous supersymmetry breaking
\cite{Burgess:2003ic}. In either case, one ends up with a
metastable dS state which can decay by tunneling and formation of
bubbles of 10d space with vanishing vacuum energy density. The
decay rate is extremely small \cite{KKLT}, so for all practical
purposes, one obtains an exponentially expanding de Sitter space
with the stabilized volume of the internal space\footnote{It is
also possible to find de Sitter solutions in noncritical string
theory \cite{Str}.}.

\subsection{Inflation in string theory}

There are two different versions of string inflation. In the first version, which we will call modular inflation, the inflaton field is  associated with one of the moduli, the scalar fields which are already present in the KKLT construction. In the second version, the inflaton is related to the distance between branes moving in the compactified space. (This scenario should not be confused with  inflation in the brane world scenario  \cite{Arkani-Hamed:1998rs,Randall:1999ee}. This is a separate interesting subject, which we are not going to discuss in this paper.)

\subsubsection {Modular inflation}

An example of the  KKLT-based  modular inflation is provided by the racetrack inflation model of Ref. \cite{Blanco-Pillado:2004ns}. It uses a slightly more complicated superpotential  
\begin{equation}\label{race}
W=W_0+
Ae^{-a\rho} + B e^{-b\rho}.
 \end{equation}
The potential of this theory has a saddle point as a function of  the real and the complex part of the volume modulus: It has a local minimum in the direction $\rm Re\, \rho$,  which is simultaneously a very flat maximum with respect to $\rm Im\,\rho$. Inflation occurs during a slow rolling of the field $\rm Im\,\rho$ away from this maximum (i.e. from the saddle point). The existence of this regime requires a significant fine-tuning of  parameters of the superpotential. However, in the context of the string landscape scenario describing from $10^{100}$ to $10^{1000}$ different vacua (see below), this may not be such a big issue. A nice feature of this model is that it does not require adding any new branes  to the original KKLT scenario, i.e. it is rather economical. Another attractive feature of this model is the existence of the regime of eternal  inflation near the saddle point.

\begin{figure}[h!]
\centering\leavevmode\epsfysize=9cm \epsfbox{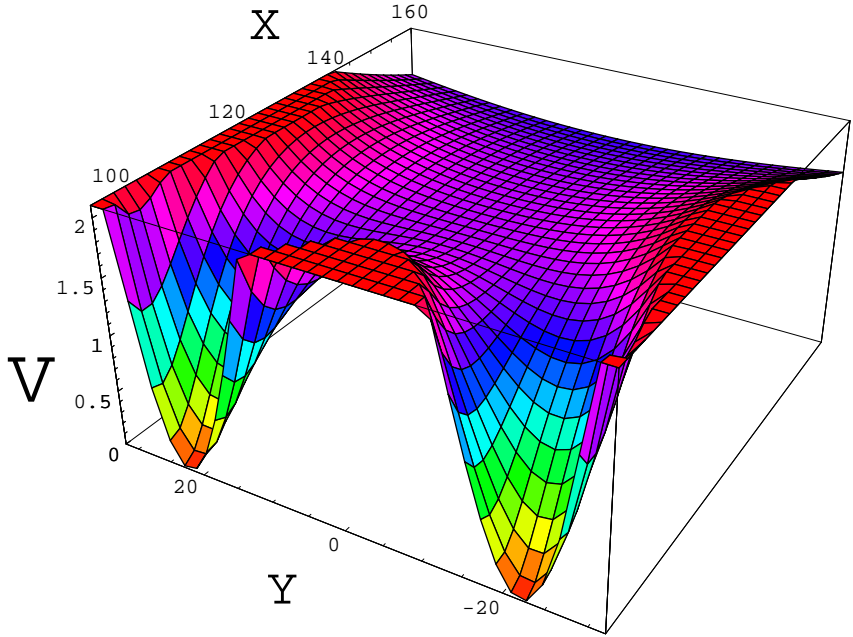}

\caption[fig1] {Plot for the  potential in the racetrack model (rescaled by
$10^{16}$). Here X stays for $\sigma = {\rm Re }\, \rho$ and Y stays for $\alpha = {\rm Im }\, \rho$. Inflation begins in a vicinity of the saddle point at $X_{\rm saddle}=123.22$, $ Y_{\rm saddle}=0$.
Units are $M_p=1$.\label{F1}}
\end{figure}

\subsubsection {Brane inflation and shift symmetry}

During the last few years there were many suggestions how to
obtain hybrid inflation in string theory by considering motion of
branes in the compactified space, see \cite{Dvali:1998pa,Quevedo}
and references therein. The main problem of all of these models
was the absence of stabilization of the compactified space. Once
this problem was solved for dS space \cite{KKLT}, one could try to
revisit these models and develop models of brane inflation
compatible with the volume stabilization.

The first idea \cite{KKLMMT} was to consider a pair of D3 and
anti-D3 branes in the warped geometry studied in \cite{KKLT}. The
role of the inflaton field  $\phi$ could be played by the interbrane
separation. A description of this situation in terms of the
effective 4d supergravity involved K\"ahler potential
\begin{equation}
K = -3\log (\rho+\bar\rho -k(\phi,\bar\phi)), \end{equation}
 where the function
$k(\phi,\bar\phi)$ for the inflaton field $\phi$, at small $\phi$,
was taken in the simplest form $k(\phi,\bar\phi)= \phi\bar\phi$.
 If one makes  the simplest
assumption that the superpotential does not depend on $\phi$, then
the $\phi$ dependence of the potential (\ref{superpot})  comes
from the term $e^K =(\rho+\bar\rho - \phi\bar\phi)^{-3}$.
Expanding this term near the  stabilization point $\rho = \rho_0$,
one finds that the inflaton field has a mass $m^2_\phi = 2H^2$.
Just like the similar relation $m^2_\phi = 3H^2$ in the simplest
models of supergravity, this is not what we want for inflation.

One way to solve this problem is to consider $\phi$-dependent
superpotentials. By doing so, one may fine-tune $m^2_\phi$ to be
$O(10^{-2}) H^2$ in a vicinity of the point where inflation occurs
\cite{KKLMMT}. Whereas fine-tuning is certainly undesirable, in
the context of string cosmology it may not be a serious drawback.
Indeed, if there exist many realizations of string theory
\cite{Douglas}, then one might argue that all realizations not
leading to inflation can be discarded, because they do not
describe a universe in which we could live. Meanwhile, those
non-generic realizations, which lead to eternal inflation,
describe inflationary universes with an indefinitely large and
ever-growing volume of inflationary domains. This makes the issue
of fine-tuning less problematic. A more detailed investigation of this issue can be found in \cite{Buchel:2003qj}.

Can we avoid fine-tuning altogether? One of the possible ideas is
to find theories with some kind of shift symmetry. Another
possibility is to construct something like D-term inflation, where
the flatness of the potential is not spoiled by the term $e^K$.
Both of these ideas were explored in Ref.
\cite{Hsu:2003cy} based on the model of D3/D7 inflation in string
theory \cite{renata}. In this model the K\"ahler potential is
given by
\begin{equation}
K = -3\log (\rho+\bar\rho) -{1\over 2}(\phi - \bar \phi)^2,
\end{equation}
and superpotential depends only on $\rho$. The role of the inflaton field is played by the field $s = {\rm Re}\, \phi$, which represents the distance between the D3 and D7 branes. The shift symmetry
$s \to s+c$ in this model is related to the requirement of
unbroken supersymmetry of branes in a BPS state.

\begin{figure}[h!]
\centering\leavevmode\epsfysize=6,6cm\hskip -0.7cm  \epsfbox{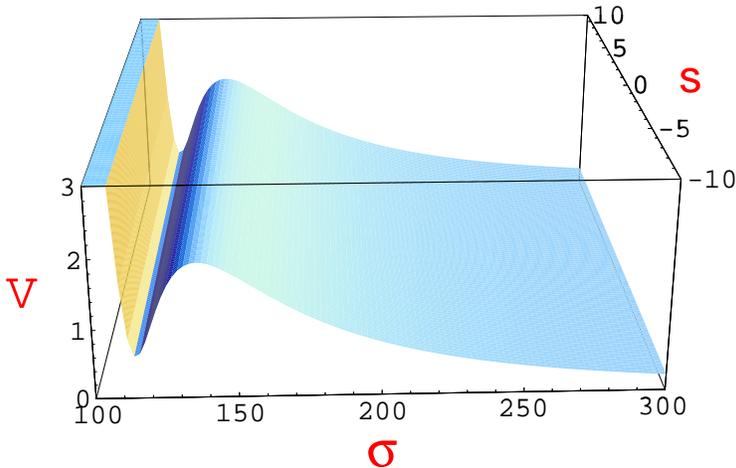} \caption[fig2]
{Inflationary potential  as a function of the compactification modulus $\sigma$ and 
the inflaton field $s$.
This potential is exactly flat in the inflaton direction due to shift symmetry, which is
violated only by radiative corrections.} \label{3}
\end{figure}

The effective potential with respect to the field $\rho$ in this
model coincides with the KKLT potential
\cite{KKLT,Burgess:2003ic}. The
potential is exactly flat in the direction of the inflaton field $s$, see Fig. \ref{3}, until one adds a hypermultiplet of  other fields $\phi_{\pm}$,  which break
this flatness due to quantum corrections and produce a logarithmic potential for the field $s$. The resulting potential with respect to the fields $s$ and $\phi_{\pm}$ is very similar to the potential of D-term hybrid inflation \cite{D}.

\begin{figure}[h!]
\centering\leavevmode\epsfysize=7.5cm \epsfbox{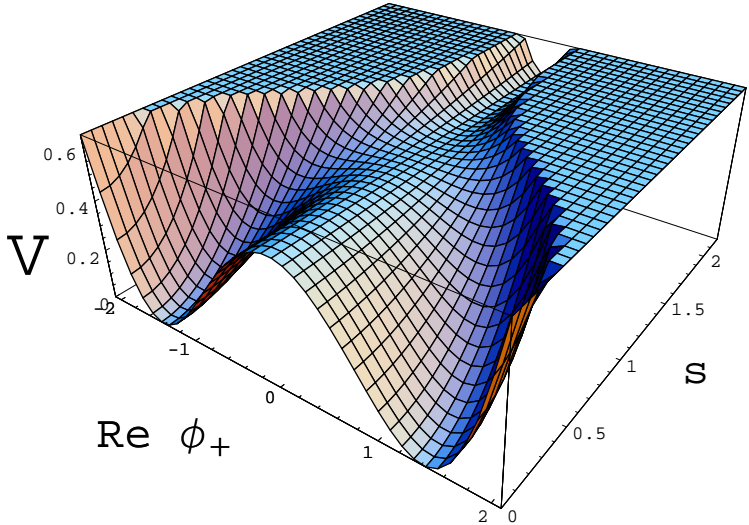} \caption[fig2]
{Inflationary potential as a function of the inflaton field $s$ and
${\rm Re}\, \phi_+$. In the beginning, the field $s$ rolls
along the valley $\phi_+=0$, and then it falls down to the KKLT minimum.}
\label{4}
\end{figure}

During inflation, $\phi_{\pm}=0$, and the field $s$ slowly rolls down to its smaller values. When it becomes sufficiently small, the theory becomes unstable with respect to generation of the field $\phi_{+}$, see Fig. \ref{4}.  The fields $s$ and  $\phi_{+}$ roll down to the KKLT minimum, and inflation ends.

One may wonder whether the shift
symmetry is just a condition which we want to impose on the theory
in order to get inflation \cite{Tye}, or an unavoidable property of the
theory, which remains valid even after the KKLT volume
stabilization. The answer to this question appears to be model-dependent. It was shown in
\cite{Angelantonj:2003up,kalrecent} that in a certain class of models, including
D3/D7 models
\cite{renata,Hsu:2003cy, Koyama:2003yc}, the
shift symmetry of the effective 4d theory is not an assumption but
an unambiguous consequence of the underlying mathematical
structure of the theory. This may allow us to obtain a natural
realization of inflation in string theory\footnote{This issue was recently debated in \cite{McAllister:2005mq}, but  the brane configuration in the model discussed there (one D3 brane interacting with a stack of many coincident D7 branes) is quite different from the configuration  considered in D3/D7 scenario of \cite{renata,Hsu:2003cy,kalrecent}.}.
For the latest developments in D3/D7 inflation see \cite{Dasgupta:2004dw,Chen:2005ae,KLupdate}.

All inflationary models discussed above were formulated in the context of Type IIB string theory with the KKLT stabilization. A discussion of the possibility to obtain inflation in the heterotic string theory with stable compactification can be found in  \cite{Buchbinder:2004nt,Becker:2005sg}.

Finally, we should mention that making the effective potential flat is not the only way to achieve inflation. There are some models with nontrivial kinetic terms where inflation may occur even without any potential \cite{kinfl}. One may also consider models with steep potentials but with anomalously large kinetic terms for the scalar fields  see e.g. \cite{Dim}. In application to string theory,  such models, called `DBI inflation,' were developed in \cite{Silverstein:2003hf}.

\section{Scale of inflation, scale of SUSY breaking and the gravitino mass}

\begin{figure}[h!]
\centering\leavevmode\epsfysize=6cm \epsfbox{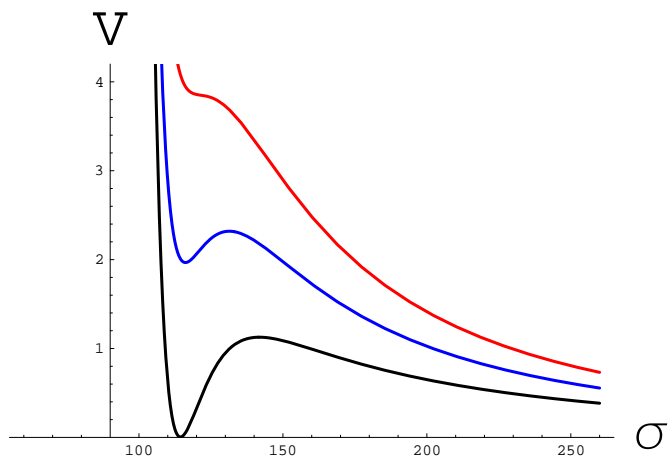} \caption[fig2]
{The lowest curve with dS minimum is the one from the KKLT model. The height of the barrier in this potential is of the order $m_{3/2}^{2}$.  The second line shows the $\sigma$-dependence of the D3/D7 inflationary potential with the term $V_{\rm infl}={V(s,\phi_{\pm})\over \sigma^3}$ added to the KKLT potential; it originates from fluxes on D7 brane. The top curve shows that when the  inflationary potential becomes too large, the barrier disappears, and the internal space decompactifies. This explains the origin of the constraint $H\lesssim   m_{3/2}$.  This constraint appears in a broad class of inflationary models based on the simplest  KKLT potential. } \label{2}
\end{figure}

So far, we did not discuss relation of the new class of models with particle phenomenology. This relation is rather unexpected and may impose strong constraints either on particle phenomenology or on inflationary models: Recently it was shown that the Hubble constant and the inflaton mass   in the simplest models based on the KKLT mechanism with the superpotential (\ref{KKLTsp}) are always smaller than the gravitino mass  \cite{Kallosh:2004yh},
\begin{equation}
H \lesssim m_{{3/2}} \ .
\end{equation}
The reason for this constraint is that adding a large vacuum energy density  to the KKLT potential may destabilize it, see Fig. \ref{2}.

Since in the slow-roll models the inflaton mass must be much smaller than $H$, its mass must be much smaller than $m_{{3/2}}$. Therefore if one insists on the standard SUSY phenomenology assuming that the gravitino mass is smaller than $O(1)$ TeV, one will need to find realistic particle physics model where  the nonperturbative string theory dynamics occurs at the LHC scale (!!!), and inflation occurs  at least 30 orders of magnitude below the Planck energy density. Such models are possible, but their parameters should be substantially different from the parameters used in all presently existing models of string theory inflation.

There are several different ways to address this problem. First of all, one may try to construct realistic particle physics models with superheavy gravitinos \cite{DeWolfe:2002nn,Arkani-Hamed:2004fb}. Another possibility is to consider models with the racetrack superpotential (\ref{race}) and find such parameters that the minimum of the potential even before the uplifting will occur at vanishingly small energy density. This goal was achieved in  \cite{Kallosh:2004yh}. 

\begin{figure}[h!]
\centering\leavevmode\epsfysize=6cm \epsfbox{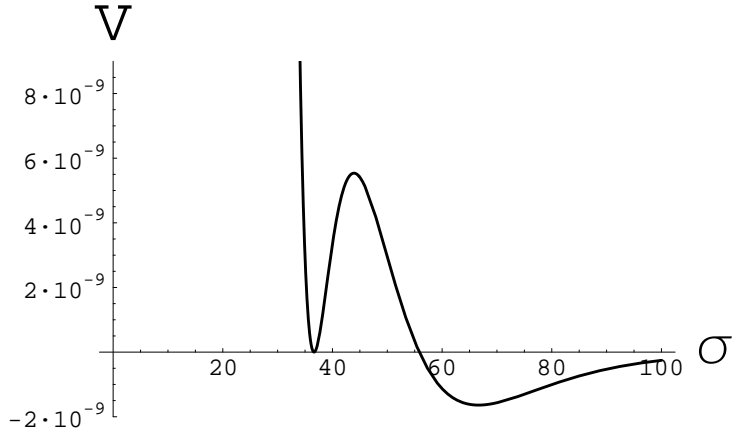} \caption[fig2]
{The potential in the theory (\ref{race}) for $A=1,\ B=-5,\ a=2\pi/100,\ b=2\pi/50,\ W_0= -0.05$. A Minkowski minimum at $V=0$ stabilizes the volume at $\sigma_{0}\approx 37$. AdS vacuum at $V<0$ occurs at $\sigma_{0}\sim 66$.  There is a barrier protecting the Minkowski minimum, with the height $V_{{\rm B}}\sim 5\times 10^{-9}$, in units of the Planck energy density. The height of the barrier is not correlated with the gravitino mass, which   vanishes  if the system is trapped in Minkowski vacuum. } \label{3a}
\end{figure}

An intriguing property of the new version of the KKLT construction is that the minimum of the potential prior to the uplifting corresponds to a supersymmetric Minkowski vacuum. The gravitino mass in this minimum (and the magnitude of SUSY breaking there) can be vanishingly small as compared to all other parameters of the model, and the constraint $H \lesssim m_{{3/2}}$ disappears. Anywhere outside this minimum the gravitino mass and the magnitude of SUSY breaking is extremely large.
This means that the Minkowski minimum shown in Fig. \ref{3a} is  a point of enhanced symmetry, which is a trapping point for the motion of the moduli fields, in accordance with \cite{Kofman:2004yc}. This fact may increase the probability that among all possible minima in the string theory landscape, the minimum with a low-scale SUSY breaking is dynamically preferred.

The original KKLT model with the potential shown by the black line in Fig. \ref{1}, and its new version with the potential shown in Fig. \ref{3a}, represent the two presently available models of dark energy based on string theory. In both cases the universe experiences constant acceleration determined by the small positive vacuum energy in the minimum of the potential. In both cases, these vacuum states are expected to be metastable, with an extremely long lifetime. These states may decay by forming expanding bubble of a new phase. It would be very hard to detect such a bubble and even harder to report the results of the observations, since the bubble walls move with the speed approaching the speed of light, so an observer would see the bubble wall at the moment when   it hits him.

The difference between these two versions is that the interior of the bubble in the simplest version of the KKLT scenario represents a 10D Minkowski space, whereas an interior of the bubble in its modified version shown  in Fig. \ref{3a} will correspond to a very rapidly collapsing open universe with a negative energy density.

The only good thing about this grim picture is that in an eternally existing self-reproducing universe there will always remain some parts of space where somebody else will enjoy life.

\section{Initial conditions for the low-scale inflation and topology of the universe}\label{torus}

One of the advantages of the simplest chaotic
inflation scenario is that inflation may begin in the universe immediately after its creation at the largest
possible energy density $M_p^{4}$, of a smallest possible size (Planck
length), with the smallest possible mass $M \sim M_p$ and with the
smallest possible entropy $S = O(1)$. This provides a true
solution to the flatness, horizon, homogeneity, mass and entropy
problems \cite{book}. 

Meanwhile, in the new inflation scenario, inflation
occurs on the mass scale 3 orders of magnitude below $M_p$, when
the total size of the universe was very large. If, for example, the universe is closed, its total mass  at the beginning of new inflation
must be  greater than $10^6 M_p$, and its total entropy must be
greater than $10^9$. In other words, in order to explain why the
entropy of the universe at present is greater than $10^{87}$ one should
assume that it was extremely large from the very beginning. This
does not look like a real solution of the entropy problem. A
similar problem exists for many of the models advocated in
\cite{Lyth:2003kp,Boubekeur:2005zm}. Finally, in cyclic inflation, the process of
exponential expansion of the universe occurs only if the total
mass of the universe is greater than its present mass $M \sim
10^{60} M_p$ and its total entropy is greater than $10^{87}$. This
scenario does not solve the flatness, mass and entropy problems.

Is it at all possible to solve the problem of initial conditions for the low scale inflation?
The answer to this question appears to be positive though perhaps somewhat unexpected: One should consider a compact flat  or open universe with nontrivial topology (usual flat or open universes are infinite). The universe may initially look like a nearly homogeneous torus of a Planckian size containing just one or two photons or gravitons. It can be shown that such a universe continues expanding and remains homogeneous until the onset of inflation, even if inflation occurs only at a very low scale \cite{ZelStar,chaotmix,topol4,Coule,Linde:2004nz}. 

Consider for simplicity the flat compact universe having the topology of a
torus, $S_1^3$,
\begin{equation}\label{2t}
ds^2 = dt^2 -a_i^2(t)\,dx_i^2
\end{equation}
with identification $x_i+1 = x_i$ for each of the three dimensions. Consider
for  simplicity the case $a_1 = a_2 = a_3 = a(t)$. In this case the curvature
of the universe and the Einstein equations written in terms of $a(t)$ will be
the same as in the infinite flat Friedmann universe with metric $ds^2 = dt^2
-a^2(t)\,d{\bf x^2}$. In our notation, the scale factor $a(t)$ is equal to the
size of the universe in Planck units $M_{p}^{{-1}} = 1$.

Let us assume,  that at the
Planck time $t_p \sim M_p^{-1}=1$ the universe was radiation dominated, $V\ll
 T^4 = O(1)$. Let us also assume that at the Planck time the total size of the
box was Planckian, $a(t_p) = O(1)$. In such case the whole universe initially
contained only $O(1)$ relativistic particles such as photons or gravitons, so
that the total entropy of the whole universe was O(1).

 The size of the universe dominated by relativistic particles was growing as
$a(t) \sim \sqrt t$, whereas the mean free path of the gravitons was growing as
$H^{-1}\sim t$. If the initial size of the universe was $O(1)$, then at the
time  $t \gg 1$ each particle (or a gravitational perturbation of metric)
within one cosmological time would run all over the torus many times, appearing
in all of its parts with nearly equal probability. This effect, called
``chaotic mixing,'' should lead to a rapid homogenization of the universe
\cite{chaotmix,topol4}. Note, that to achieve a modest degree of homogeneity
required for inflation to start when the density of ordinary matter drops down,
we do not even need chaotic mixing. Indeed, density perturbations do not grow
in a universe dominated by ultrarelativistic particles if  the size of the
universe is smaller than $H^{-1}$. This is exactly what
happens in our model. Therefore the universe should remain relatively
homogeneous until the thermal energy density drops below $V$ and inflation
begins.

Thus we see that in this scenario, just as in the simplest chaotic inflation scenario, inflation begins if we had a sufficiently homogeneous domain of a smallest possible size (Planck scale), with the smallest possible mass (Planck mass), and with the total entropy O(1). The only additional requirement is that this domain should have identified sides, in order to make a flat or open universe compact. We see no reason to expect that the probability of formation of such domains is strongly suppressed.

One can come to a similar conclusion from a completely different point of view. Investigation of the quantum creation of a closed or  an infinite open inflationary universe with a small value of the vacuum energy shows that this process is forbidden at the classical level, and therefore it occurs only due to tunneling. As a result, the probability of this process is exponentially suppressed \cite{Linde:1983mx,Vilenkin:1984wp,Open}. Meanwhile,  creation of the flat or open universe  is possible without any need for the tunneling, and therefore there is no exponential suppression for the probability of quantum creation of a topologically nontrivial compact flat or open inflationary universe \cite{ZelStar,Coule,Linde:2004nz,KLS2005}.

These results suggest that if inflation can occur only  much below the Planck density, then the topologically nontrivial flat or open universes should be much more probable than the standard Friedmann universes described in every textbook on cosmology.\footnote{Note, however, that unless one fine-tunes the parameters of the theory and the shape of the inflationary potential, inflation  makes the observable part of the universe so large that its topology should not affect observational data.}

Until now, we discussed creation of  compact universes which have approximately equal size in all directions.  If at the Planck  time our universe was of a Planck size in some  directions, but it was much larger in some other directions, then it consisted of many causally disconnected Planck size regions. This results in a particular version of the horizon/homogeneity problem: The probability that the  universe was homogeneous in all of these causally disconnected regions should be exponentially small \cite{KLS2005}. 

In application to the initial conditions in the 10D universe described by string theory, this suggests that it is more natural to start with the universe which would have similar initial size in all 9 spatial directions. In terms of the KKLT potential, this implies that the initial value of the volume modulus $\sigma$ should be very small, so that $V(\sigma) =O(1)$. But then the field $\sigma$ will rapidly fall down. It can easily roll over the very  law KKLT barrier, and continue moving to infinitely large $\sigma$. 

One of the recent attempts to solve this problem was based on the dynamical slow-down of the field $\sigma$ in the universe filled with radiation \cite{Brustein:2004jp}.
But this mechanism typically works only if the initial value of the effective potential is several orders of magnitude smaller than O(1). 

It is not quite clear whether this is a real problem   because those parts of the universe  which start at large $V(\sigma)$, become ten-dimensional, so  we cannot live there. We will live in the parts of the universe which started at smaller $V(\sigma)$, even if it may seem improbable from the point of view of initial conditions. This is  similar to the fact that we live at the 2D surface of a relatively small planet instead of living in the vast but empty  interstellar space.

One can also argue  \cite{LLM} that eternal inflation may alleviate
some of the problems of the problem of initial conditions for the low-scale inflation. Note, however,
that eternal inflation, which naturally occurs in the simplest
versions of chaotic inflation,  in new inflation, and in the racetrack scenario \cite{Blanco-Pillado:2004ns}, may not exist in many  versions of string theory  inflation of the hybrid type, such as the models of \cite{KKLMMT,Hsu:2003cy,KLupdate}. Of course,  models
of low-scale non-eternal inflation are still much better than the
models with no inflation at all, but I do not think that we should
settle for the second-best. An indeed we should not, if we are able to combine the slow-roll inflation, which ends in our vacuum, with the eternal inflation in a collection of different metastable dS states in the string theory landscape.

\section{Eternal inflation and the string theory landscape}\label{land}

Even though we are still at the very first stages of implementing
inflation in string theory, it is very tempting to speculate about
possible generic features and consequences of such a construction.

First of all, KKLT construction shows that the vacuum energy after
the volume stabilization is a function of many different
parameters in the theory. One may wonder how many different
choices do we actually have. There were many attempts to
investigate this issue. For example, many years ago it was argued
\cite{Duff} that there are infinitely many $AdS_4 \times X7$ vacua
of D=11 supergravity. An early estimate of the total number of
different 4d string vacua gave the number $10^{1500}$
\cite{Lerche}. At present we are more interested in counting
different flux vacua \cite{BP,Douglas}, where the possible
numbers, depending on specific assumptions, may vary in the range
from $10^{20}$ to $10^{1000}$. Some of these vacuum states with
positive vacuum energy can be stabilized using the KKLT approach.
Each of such states will correspond to a metastable vacuum state.
It decays within a cosmologically large time, which is, however,
smaller than  the `recurrence time' $e^{S(\phi)} $, where $S(\phi)
= {24\pi^2\over V(\phi)}$ is the entropy of dS space with the
vacuum energy density $V(\phi)$ \cite{KKLT}.

But this is not the whole story; old inflation does not describe
our world. In addition to these metastable vacuum states, there
should exist various slow-roll inflationary solutions, where the
properties of the system practically do not change during the
cosmological time $H^{-1}$. It might happen that such states,
corresponding to flat directions in the string theory landscape,
exist not only during inflation in the very early universe, but
also at the present stage of the accelerated expansion of the
universe. This would simplify obtaining an anthropic solution of
the cosmological constant problem along the lines of
\cite{Linde84,BP}.

If the slow-roll condition $V'' \ll V$ is satisfied all the way
from one dS minimum of the effective potential to another, then
one can show, using stochastic approach to inflation, that the
probability to find the field $\phi$ at any of these minima, or at
any given point between them,  is proportional to $e^{S(\phi)}$.
In other words, the relative probability to find the field taking
some value $\phi_1$ as compared to some other value $\phi_0$, is
proportional to $e^{\Delta S} = e^{S(\phi_1)-S(\phi_0)}$
\cite{Open,KKLT}. One may argue, using Euclidean approach, that
this simple thermodynamic relation should remain valid for the
relative probability to find a given point in any of the
metastable dS vacua, even if the trajectory between them does not
satisfy the slow-roll condition $m^2 \ll H^2$
\cite{HM,Lee:qc,Garriga:1997ef,Dyson:2002pf}.

The resulting picture resembles eternal inflation in the old
inflation scenario. However, now we have an incredibly large
number of false vacuum states, plus some states which may allow
slow-roll inflation. Once inflation begins, different parts of the
universe start jumping from one of these vacuum states to
another, so that the universe becomes divided into indefinitely
many regions with all possible laws of low-energy physics
corresponding to different 4D vacua of string theory \cite{book}.

The best inflationary scenario would
describe a slow-roll eternal inflation starting at the maximal
possible energy density (minimal dS entropy). It would be almost
as good to have a low-energy slow-roll eternal inflation. Under
certain conditions, such regimes may exist in string theory
\cite{KKLMMT}. Here we are going to discuss a different but equally interesting possibility.
Multi-level eternal inflation of the old inflation type, which
appears in string theory in the context of the KKLT construction,
may be very useful being combined with the slow-roll inflation,
even if the slow-roll inflation by itself is not eternal. We will
give a particular example, which is based on the ideas developed in Ref. \cite{Linde:1987yb}.

Suppose we have two noninteracting scalar fields: field $\phi$
with the potential of the old inflation type, and field $s$
with the potential which may lead to a slow-roll inflation. Let us
consider the worst case
scenario:  the slow-roll inflation  occurs only on low
energy scale, and it is not eternal.  

Let us assume that the Hubble constant at the stage of old
inflation is much greater than the curvature of the potential
which drives the slow-roll inflation. This is a natural
assumption, considering huge number of possible dS states, and the
presumed smallness of energy scale of the slow-roll inflation. The validity of this assumption is  ensured if the shift symmetry with respect to the slow-rolling inflaton field $s$ is  preserved in many different string theory vacua. This may be the case in the D3/D7 scenario, where the shift symmetry is related to the isometry of the compactified space but not to the particular values of the fluxes  \cite{Hsu:2003cy,kalrecent,KLupdate}. In this case large inflationary fluctuations of the field $s$ will
be generated during eternal old inflation. These fluctuations will push the field $s$ from the minimum of its effective potential and will 
give it different values in different exponentially
large parts of the universe. When old inflation ends, there will
be many practically homogeneous parts of the universe where  the
field $s$ will take  values corresponding to good initial
conditions for a slow-roll inflation.  

Thus, the existence of many different dS vacua in string theory
leads to the regime of eternal inflation. This regime may help us
to solve the problem of initial conditions for the slow-roll
inflation even in the models where the slow-roll inflation by
itself is not eternal and would occur only on a small energy
scale.

 \begin{acknowledgments}
I am grateful to the organizers of the  SLAC Summer School ``Nature's Greatest Puzzles," of the conference Cosmo04 in Toronto, of the VI Mexican School on Gravitation, and of the XXII Texas Symposium on Relativistic Astrophysics. I would like to thank Lev Kofman,  Renata Kallosh and Slava Mukhanov for  valuable comments.  This work
 was supported by NSF grant PHY-0244728.  
 \end{acknowledgments}
 

\end{document}